\documentclass[10pt,twocolumn,twoside]{IEEEtranTCOM}

\ifCLASSINFOpdf
\else
\fi
\usepackage{cite}
\usepackage{graphicx}
\usepackage{amsmath}
\usepackage{algpseudocode}
\usepackage{amssymb}
\usepackage{times}
\usepackage{endnotes}
\usepackage{stfloats}
\usepackage{stkernel}
\usepackage{cite}
\usepackage{enumerate}
\usepackage{array}
\usepackage{relsize}
\usepackage[belowskip=-10pt,aboveskip=3pt,footnotesize]{caption}

\renewcommand{\baselinestretch}{0.98}

\newcommand{\refeq}[1]{(\ref{#1})}

\title{Soft metrics and their Performance Analysis for Optimal Data Detection in the Presence of Strong Oscillator Phase Noise}

\author{Rajet~Krishnan,~\IEEEmembership{Student~Member,~IEEE,}
        M.~Reza~Khanzadi,~\IEEEmembership{Student~Member,~IEEE,}
        Thomas~Eriksson,
        and~Tommy~Svensson, ~\IEEEmembership{Member,~IEEE}

\thanks{Rajet Krishnan,  M. Reza Khanzadi, Thomas Eriksson, and Tommy Svensson are with the Department
of Signals and Systems, Chalmers Univeristy of Technology, Gothenburg, Sweden ( E-mail: \{rajet, khanzadi, thomase, tommy.svensson\}@chalmers.se).}%

}%

\begin{document}

\markboth{IEEE Transactions on Communications}%
{Accepted for Publication}

\maketitle \thispagestyle{empty}

\begin{abstract}
In this paper, we address the classical problem of maximum-likelihood (ML) detection of data in the presence of random phase noise. We consider a system, where the random phase noise affecting the received signal is first compensated by a tracker/estimator. Then the phase error and its statistics are used for deriving the ML detector. Specifically, we derive an ML detector based on a Gaussian assumption for the phase error probability density function (PDF). Further without making any assumptions on the phase error PDF, we show that the actual ML detector can be reformulated as a weighted sum of central moments of the phase error PDF. We present a simple approximation of this new ML rule assuming that the phase error distribution is unknown.  The ML detectors derived are also the aposteriori probabilities of the transmitted symbols, and are referred to as soft metrics. Then, using the detector developed based on Gaussian phase error assumption, we derive the symbol error probability (SEP) performance and error floor analytically for arbitrary constellations. Finally we compare SEP performance of the various  detectors/metrics  in this work and those from literature for different signal constellations, phase noise scenarios and SNR values.

\emph{Index Terms} - Maximum likelihood detection, Estimation, Phase noise.


\end{abstract}

\section{Introduction}
\label{sec:intro}

Random time varying phase difference between the transmitter and receiver has been one of the major impediments towards realizing a reliable coherent communication system \cite{marc99}. This impairment, which arises from local oscillator instabilities, is referred to as phase noise, and can result in significant performance loss if not compensated appropriately. Given the demands for high-rate data transmission over band-limited channels, high order signal constellations are being considered for transmission, where even minor phase noise impairments can incur heavy loss in system performance.

The problem of receiver design for data detection in the presence of a random phase noise process has been studied for decades, e.g., refer to \cite{marc99,meng97} and references therein. One of the earlier approaches to this problem was reported in \cite{koba71}, which proposed simultaneous \textit{maximum-likelihood} (ML) estimation of the data sequence and phase noise. In \cite{falc77}, it was shown that the simultaneous approach proposed in \cite{koba71} is optimal in the high \textit{signal-to-noise ratio} (SNR) regime.

An optimum minimum SEP criterion receiver structure was first derived by Kam \textit{et al.} in \cite{kam94}. Specifically, it was illustrated that the optimum \textit{symbol-by-symbol} (SBS) receiver has a separable estimator-detector structure, i.e., all the received signals (or observations) are used to first compute/estimate the posteriori \textit{probability density function} (PDF) of phase noise. The information in this posteriori phase density function is then used to detect a data symbol. Further the problem of computing the posteriori PDF of phase noise given all the received signals was demonstrated to be an intractable problem in general. However, it was seen that the optimum receiver structure can be analytically determined under restrictive assumptions on the phase noise distribution.

On a related note, the idea of using a tractable phase posteriori density for deriving the ML detector for data was reported in a much earlier work by Foschini \textit{et al.} in \cite{fos73}. It was assumed that the phase of the received signal is first tracked and compensated using a \textit{phase-locked loop} (PLL) or an estimator. Then the posteriori phase PDF, which now is the PDF of the \textit{phase error}, can be approximated as a Tikhonov PDF \cite{vit63} which was used to derive the ML detector. In a more recent effort, an ML detector similar to that in \cite{fos73} was derived in \cite{imai04} for phase noise limited channels, and used as a \textit{soft metric} for decoding in coded systems.

In recent times, there has been great interest in understanding the performance of coded systems in the presence of random phase noise impairments. In these scenarios, techniques like per-survivor processing algorithm proposed in \cite{rahe95}, and turbo synchronization methods based on \text{expectation maximization} (EM) reported in \cite{lott04,anas01,noel05,noel07,ferr07} have been extensively used for joint (and iterative) phase estimation and data detection. In these approaches, the phase perturbance of the transmitted signal is regarded as a deterministic constant that is first estimated. Treating the estimate as the true value of phase perturbance, the detector computes aposteriori symbol probabilities (depending on the underlying code structure) assuming that the received signal is devoid of any phase noise. Broadly speaking, these techniques can be considered as non-Bayesian approaches where the carrier phase noise is considered deterministic and constant over a block of symbols transmitted. Clearly, these techniques are inappropriate for scenarios where the phase noise is random and time varying.

To address the problem of random phase perturbance of the transmitted symbols, iterative Bayesian techniques based on factor graph signal processing for data detection and phase noise estimation were proposed in \cite{dauw03,dauw04,cola05}. These techniques that are based on the generic \textit{sum-product algorithm} (SPA) \cite{loe01} do not employ an explicit estimator, and are used to derive approximate Bayesian detectors based on marginalization of phase noise. As a low complexity alternative to SPA, the work in \cite{vb09} proposed joint phase estimation and approximate Bayesian detection based on the \textit{Variational-Bayesian} (VB) framework, which was found to be efficient in the presence of random phase noise. Application of Monte Carlo sampling methods for joint phase noise estimation and data detection was investigated in \cite{marc09} for both coded and uncoded systems. Though joint phase-data estimators have been studied extensively, it is not well understood as to how phase error and its statistics can be incorporated in the detector for computing aposteriori symbol probabilities \cite{noel07}.

For signal constellations with equiprobable symbols, the ML detection rule renders aposteriori symbol probabilities, which can be used as soft metric input for performing non-data aided phase estimation and detection \cite{noel05}. Importantly, soft metrics can also be embedded into iterative decoding schemes for coded systems in the presence of phase noise \cite{imai04}, \cite{cola05}, \cite{vb09} that can significantly enhance SEP performance. The ML detector derived by Foschini \textit{et al.} \cite{fos73} considering phase error PDF is one of the few detectors available in literature for detection in the presence of random phase noise. However, this detector is difficult to analyze and its SEP performance is computable only numerically \cite{kam09}.

This motivates the need to derive detectors that have simpler analytical structures, and whose performance is easier to characterize analytically. Further as we shall observe in the sequel, it is interesting to derive detectors without making any restrictive assumption on the PDF of the phase error and analyze their performance. These aspects form the core of this work.

\subsection{Contributions}
In this paper\footnote{The material in this paper was presented in part at the Global Telecommunications Conference (GLOBECOM 2011), Houston, TX, USA}, motivated by the optimal receiver structure derived in \cite{kam94} and \cite{fos73}, we revisit the fundamental problem of optimal data detection in the presence of random phase noise. We seek to minimize SEP, which is achieved by SBS detection, based on the entire set of received signals. We derive approximate analytical forms of the ML detector for different, but broad assumptions on the phase error PDF. Specifically, the phase noise is first compensated by using a tracker/estimator. The phase error (after tracking/estimation) is considered to be independent of the symbol transmitted and the other observations outside the current (received) signal interval \cite{fos73}, and its statistics are incorporated in deriving the ML detection rule.

The contributions of this work can be summarized as follows:
\begin{itemize}
\item Based on a Gaussian assumption for the phase error PDF, we derive an ML detector that corresponds to joint amplitude and phase detection of the received signals.
\item Further, without making any assumptions on the PDF of the phase error, we formulate the ML detector originally derived in \cite{kam94} as a weighted sum of the central moments of the phase error PDF. Based on the assumption that the phase error distribution is unknown, we present approximations of the new ML rule, and investigate its scope and applicability.
\item Using the detector based on the Gaussian PDF assumption for phase error, we present analytical results for its SEP performance and derive error floors for arbitrary constellations. We show that the analytical SEP results can be used to accurately determine the performance of the Gaussian based detector for different constellations, SNR values and phase noise scenarios. Interestingly, these results also provide simple insights into designing constellations that help minimize SEP in the presence of phase noise.
\item Finally, we compare the performance of all detectors proposed in this work with those existing in literature. Specifically, we show that the Gaussian based detector proposed in this work outperforms all other detectors for different constellations, SNR values and phase noise scenarios. We also use these detectors/metrics as soft input for non-data aided phase estimation and their SEP performance is studied.
\end{itemize}

Notations: Expectation operator is denoted as $\mathbb{E}[\cdot]$. $[\cdot]^{T}$ denotes transpose and $[\cdot]^{H}$ denotes Hermitian of a vector. $\text{Re}(\cdot)$, $\text{Im}(\cdot)$, $|\cdot|$ and $\text{arg}(\cdot)$ are the real, imaginary part, magnitude and angle of a complex number respectively.

\subsection{Organization }
The remainder of the paper is organized as follows: In Section \ref{sec:system_model}, the system model for detection of data in the presence of phase noise and additive white Gaussian noise (AWGN) is presented. In Section \ref{sec:analysis}, we first review the ML detectors derived in \cite{fos73}, \cite{noel05}, \cite{cola05} and \cite{vb09}. Then, we derive the ML detectors for two different assumptions on the phase error PDF. In Section \ref{sec:sep_ana}, we derive the SEP performance and error floor of the detector that is based on the Gaussian assumption for phase error. We compare the SEP performance of the various detectors for different constellations, SNR values and phase noise scenarios in Section \ref{sec:results}. Finally, we conclude our work and highlight key findings in Section \ref{sec:conc}.

\section{System Model}
\label{sec:system_model}

Consider a system with the following received signal model in the $k$th time slot
\begin{eqnarray}\label{eq:sig_mod}
r_{k}^{\prime \prime} = m_{k}e^{j\phi_{k}} + n_{k}^{\prime \prime},
\end{eqnarray}
where $r_{k}^{\prime \prime}$ is the received signal, $m_{k}$ is the transmitted symbol, $\phi_{k}$ is the unknown phase noise, and $n_{k}^{\prime \prime}$ is complex Gaussian noise in the $k$th time slot. $\textbf{r}^{\prime \prime} \triangleq \left[r_{0}^{\prime \prime},\ldots,r_{L-1}^{\prime \prime}\right]^{T}$ represents a vector of $L$ received symbols. The transmitted data are denoted in the vector form as $\textbf{m} \triangleq \left[m_{0},\ldots,m_{L-1}\right]^{T}$. 
In addition, $m_{i}$, for $i=0,\ldots,L-1$ can assume any point $\{s_{i},\, \forall \, i \in \{1,...,C\}\}$ in the signal constellation, where $C$ is the size of the constellation. Let \textbf{${\boldsymbol{\phi}}$}$ \triangleq \left[\phi_{0},\ldots,\phi_{L-1}\right]^{T}$ denote the vector of unknown phase noise, where no assumptions are made on its PDF. It is assumed that $\textbf{m}$ and $\boldsymbol{\phi}$ are independent of each other. The \textit{additive white Gaussian noise} (AWGN) is $\textbf{n}^{\prime \prime} \triangleq \left[n_{0}^{\prime \prime},\ldots,n_{L-1}^{\prime \prime}\right]^{T}$, i.e., it is a vector of \textit{independent identically distributed} (i.i.d.) complex Gaussian \textit{random variables}  (r.v.s) with $\mathbb{E}[{\textbf{n}^{\prime \prime}}] = [0,\ldots,0]^{T}$, and $\mathbb{E}[{\textbf{n}^{{\prime \prime}}}{\textbf{n}^{{\prime \prime}}}^{H}] = N_{0}\textbf{I}$, i.e., $n_{k}^{\prime \prime} \sim \mathcal{CN}(0,N_{0})$.

We investigate the problem of symbol detection based on all received signals, $\textbf{r}^{\prime\prime}$, such that the SEP is minimized. It is known that SBS detection of the $k$th symbol that minimizes SEP is obtained by ML detection \cite{vantrees}. Thus, the optimum receiver for the $k$th symbol is given by
\begin{equation}\label{eq:ml_eq}
\underset{i\in \{1,\ldots,C\}}{\operatorname{argmax}}  L_{i}(k) \triangleq \underset{i\in \{1,\ldots,C\}}{\operatorname{argmax}} p(\textbf{r}^{\prime \prime}|m_{k} = s_{i}).
\end{equation}
In \cite{kam94}, it has been shown that in the presence of phase noise, $L_{i}(k)$ reduces to the following
\begin{equation}\label{eq:orig_ml}
L_{i}(k) = \int_{-\pi}^{\pi}p(r_{k}^{\prime \prime}|m_{k} = s_{i},\phi_{k})p(\phi_{k}|\overline{\textbf{r}^{\prime \prime}}_{k},m_{k} = s_{i})d\phi_{k},
\end{equation}
where $\overline{\textbf{r}}_{k}^{\prime \prime} \triangleq \left[r_{0}^{\prime \prime},\ldots,r_{k-1}^{\prime \prime},r_{k+1}^{\prime \prime},\ldots,r_{L-1}^{\prime \prime}\right]^{T}$ refers to all signals received outside the $k$th time instant. The ML detector first involves the estimation of the posteriori PDF of phase noise in a time instant using all signals received outside it. This posteriori PDF is then used to perform data detection using \refeq{eq:orig_ml}.

In general, deriving  the ML detector in \refeq{eq:orig_ml} is a difficult problem. This is because it requires accurate knowledge of the posteriori phase PDF $p(\phi_{k}|\overline{\textbf{r}}_{k}^{\prime \prime},m_{k} = s_{i})$. The work in \cite{kam94} shows that the problem of deriving the posteriori phase PDF is an infinite dimensional problem, and its closed-form is not analytically tractable. However, there are scenarios where the posteriori phase PDF will have an approximate tractable form. Let the phase of the received signal be tracked and compensated using a tracker/estimator as follows
\begin{eqnarray}\label{eq:sig_mod_new}
r_{k} \triangleq r_{k}^{\prime\prime}e^{-j\hat{\phi}_{k}} = m_{k}e^{j\theta_{k}} + n_{k},\\
\theta_{k} \triangleq \phi_{k} - \hat{\phi}_{k}, n_{k} \triangleq n_{k}^{\prime\prime}e^{-j\hat{\phi}_{k}} \nonumber
\end{eqnarray}
where $\theta_{k}$ is the phase error, and $\hat{\phi}_{k}$ is the phase estimate. Then the optimum detector, based on the compensated received signal is
\begin{equation}\label{eq:mod_ml}
L_{i}(k) = \int_{-\pi}^{\pi}p(r_{k}|m_{k} = s_{i},\theta_{k})p(\theta_{k}|\overline{\textbf{r}}_{k},m_{k} = s_{i})d\theta_{k} ,
\end{equation}
where $p(\theta_{k}|\overline{\textbf{r}}_{k},m_{k} = s_{i})$ is the posteriori PDF of phase error $\theta_{k}$, and $\overline{\textbf{r}}_{k} \triangleq \left[r_{0},\ldots,r_{k-1},r_{k+1},\ldots,r_{L-1}\right]^{T}$ refers to all signals received outside the $k$th time instant. The phase error PDF is then constrained to be a  known PDF \cite{vit63}, thereby rendering an approximate (but tractable) ML detector as we shall see in the sequel.


\section{Detection Metrics and Analysis}
\label{sec:analysis}

In this section, we first review the detectors that were derived in \cite{fos73}, \cite{noel05}, \cite{cola05} and \cite{vb09}, and discuss their underlying assumptions. Then, we derive new analytical forms of the detector in \refeq{eq:orig_ml} and \refeq{eq:mod_ml} for two different assumptions on the phase error PDF:
\begin{itemize}
\item The phase error PDF is considered to be Gaussian distributed.
\item The phase error PDF is unknown, but its central moments are known upto a certain order.
\end{itemize}
Particularly of interest are detectors that are tractable in their exact or approximate form. The constellations considered for transmission have equally likely symbol points, and hence the ML detector renders aposteriori symbol probabilities based on observations. These aposteriori symbol probabilities (equivalently ML detectors) are referred to as \textit{soft metrics}.

\subsection{Metric based on Euclidean Distance Measure (EUC-D)}
The detector in \refeq{eq:orig_ml} reduces to a Euclidean distance detector when the carrier phase recovered by a phase estimator  $\hat{\phi}_{k}$ is treated as the true value of $\phi_{k}$, followed by coherent detection of the symbols. To illustrate this, consider the posteriori PDF of $\phi_{k}$ to be a distribution with variance zero or equivalently a delta function, i.e., $p(\phi_{k}|{\overline{\textbf{r}}_{k}}^{\prime \prime} ,m_{k} = s_{i}) = \delta(\phi - \hat{\phi}_{k})$. The ML decision rule is then derived from \refeq{eq:ml_eq} as follows
\begin{eqnarray}\label{eq:coh_det_delta}
\underset{i\in \{1,\ldots,C\}}{\operatorname{argmax}} L_{i}(k) &=& \underset{i\in \{1,\ldots,C\}}{\operatorname{argmax}} -{|r_{k}^{\prime \prime} - s_{i}e^{j\hat{\phi}_{k}}|^2}, \\
&=& \underset{i\in \{1,\ldots,C\}}{\operatorname{argmax}} -{|r_{k} - s_{i}|^2}, \nonumber
\end{eqnarray}
Thus, if the recovered $\hat{\phi}_{k}$ is treated as the true value of ${\phi}_{k}$, the ML decision rule in \refeq{eq:orig_ml} becomes equivalent to the Euclidean distance detector \cite{noel05}, \cite{proakis}. For future reference, we denote this detector as \emph{EUC-D}.

\subsection{Metric based on Tikhonov PDF assumption for Phase Error (FOS-D)}
As discussed in Section \ref{sec:system_model}, let the phase of the received signal be tracked and compensated using a PLL, i.e., $\theta_{k} = \phi_{k} - \hat{\phi}_{k}$, then the posteriori PDF of $\theta_{k}$ can be approximated as a Tikhonov PDF \cite{vit63}. Note that the Tikhonov PDF is also the entropy maximizing PDF of the circular r.v. $\theta_{k}$. Based on this assumption, an analytically tractable detector was derived by Foschini \textit{et al.} for high SNR scenarios in \cite{fos73} by using the model in \refeq{eq:sig_mod_new} and \refeq{eq:mod_ml}. Specifically it was shown that the detector, based on a Tikhonov assumption for the phase error PDF, has the form
\begin{eqnarray}\label{eq:ml_fosc}
 && \underset{i\in \{1,\ldots,C\}}{\operatorname{argmax}}\, L_{i}(k) = \underset{i\in \{1,\ldots,C\}} {\operatorname{argmax}}\, -\frac{s_{i}s_{i}^{*}}{2} + v_{i},\\
&& v_{i} \triangleq \sqrt{\left(\text{Re}\{r_{k}s_{i}^{*}\} + \frac{N_{0}}{\sigma_{\text{p}}^{2}} \right)^{2} + \left(\text{Im}\{r_{k}s_{i}^{*}\}\right)^{2}},\nonumber
\end{eqnarray}
where $\sigma_{\text{p}}^{2}$ is the variance of phase error $\theta_{k}$. The variance $\sigma_{\text{p}}^{2}$ from the PLL is a function of $\overline{\textbf{r}}_{k}$ and $m_{k}$, and is known to the detector. We refer to this detector as \emph{FOS-D}. 

\subsection{Metric based on the Factor Graph framework (COL-D)}

In \cite{cola05}, Colavolpe \textit{et al.} proposed a joint phase estimation-data detection technique based on the factor graph framework. The messages propagated on the graph using SPA were approximated as a family of Tikhonov PDFs. Based on this technique, the aposteriori symbol probabilities or equivalently the ML detector is derived using the system model in \refeq{eq:sig_mod} as
\begin{eqnarray}\label{eq:ml_col}
 && \underset{i\in \{1,\ldots,C\}}{\operatorname{argmax}}\, L_{i}(k) = \underset{i\in \{1,\ldots,C\}} {\operatorname{argmax}}\, P_{u}(s_{i}), \\
&& P_{u}(s_{i}) \triangleq \exp \left\{ \frac{-s_{i}s_{i}^{*}}{N_{0}} \right\}I_{0}\left( \left|a_{f,k} + a_{b,k} + \frac{2r_{k}^{\prime\prime}s_{i}^{*}}{N_{0}} \right|\right).\nonumber
\end{eqnarray}
Here $\text{arg}\{a_{f,k}\}$ and $\text{arg}\{a_{b,k}\}$ are the phase noise hypotheses computed in the forward and backward filter recursions respectively using SPA. This detector is denoted as \textit{COL-D}.

\subsection{Metric based on the VB framework (VB-D)}

In \cite{vb09}, the VB framework was used to derive aposteriori symbol probabilities based on the assumption that given the observation, the transmitted symbols and phase noise are independent of each other. If $\sigma_{\text{p}}^{2}$ denotes the variance of the phase error $\theta_{k}$ (obtained after phase noise compensation), using \refeq{eq:sig_mod_new} the detector is of the form
\begin{align}\label{eq:ml_vb}
 \underset{i\in \{1,\ldots,C\}}{\operatorname{argmax}}\, L_{i}(k) = \underset{i\in \{1,\ldots,C\}} {\operatorname{argmax}}\, -|r_{k} - s_{i}|^2 + \frac{s_{i}s_{i}^{*}}{2}\sigma_{\text{p}}^{2}.
\end{align}
No specific assumptions are made on the posteriori phase PDF for deriving \refeq{eq:ml_vb}, and we denote this detector as \emph{VB-D}.

\subsection{Metric based on Gaussian PDF Assumption for Phase Error (GAP-D)}

\begin{figure*}[!ht]
\begin{center}
\begin{tabular}{ccc}
\includegraphics[width = 2.1in, keepaspectratio=true]{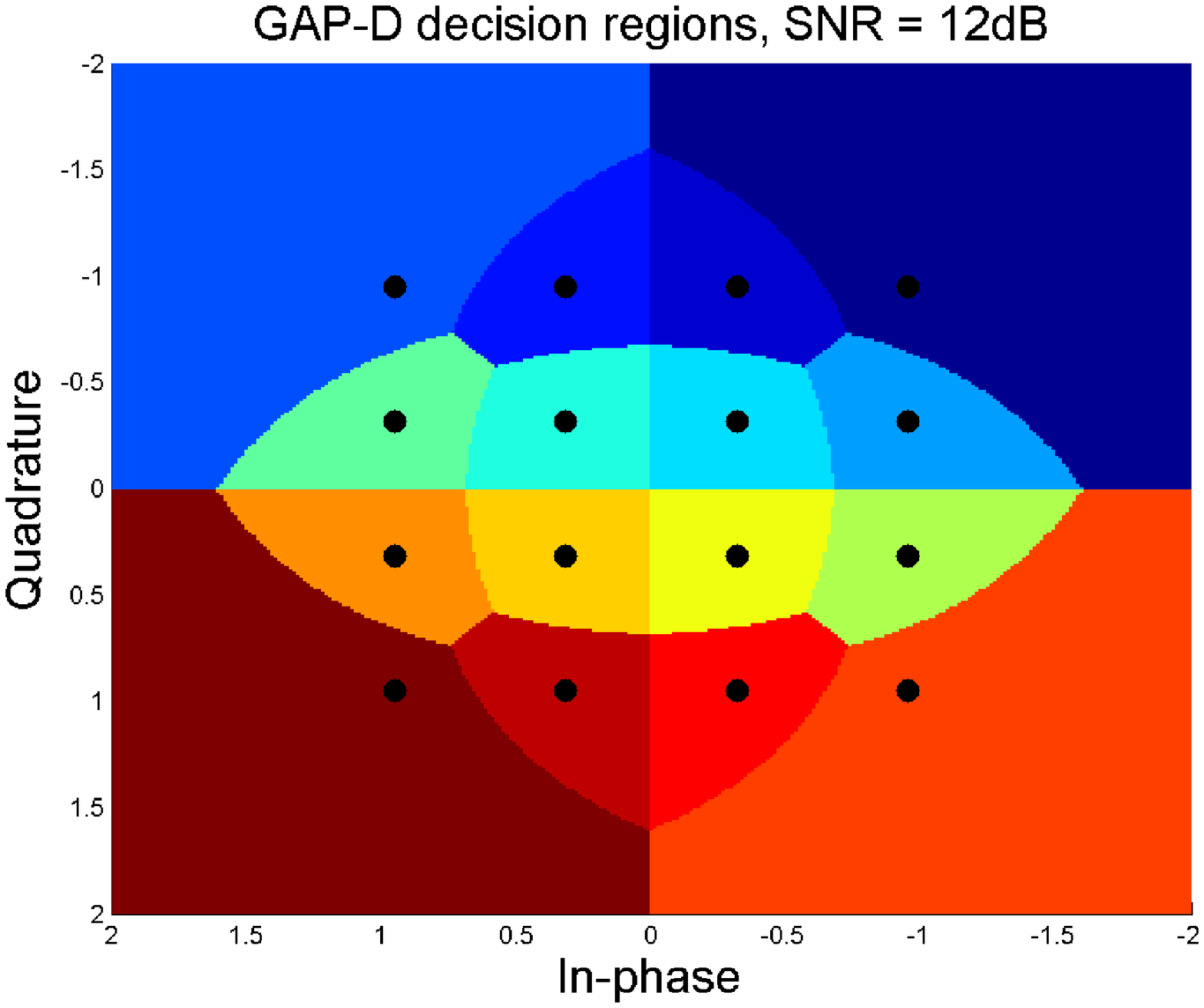} \label{fig:ML_5}
&
\includegraphics[width = 2.1in, keepaspectratio=true]{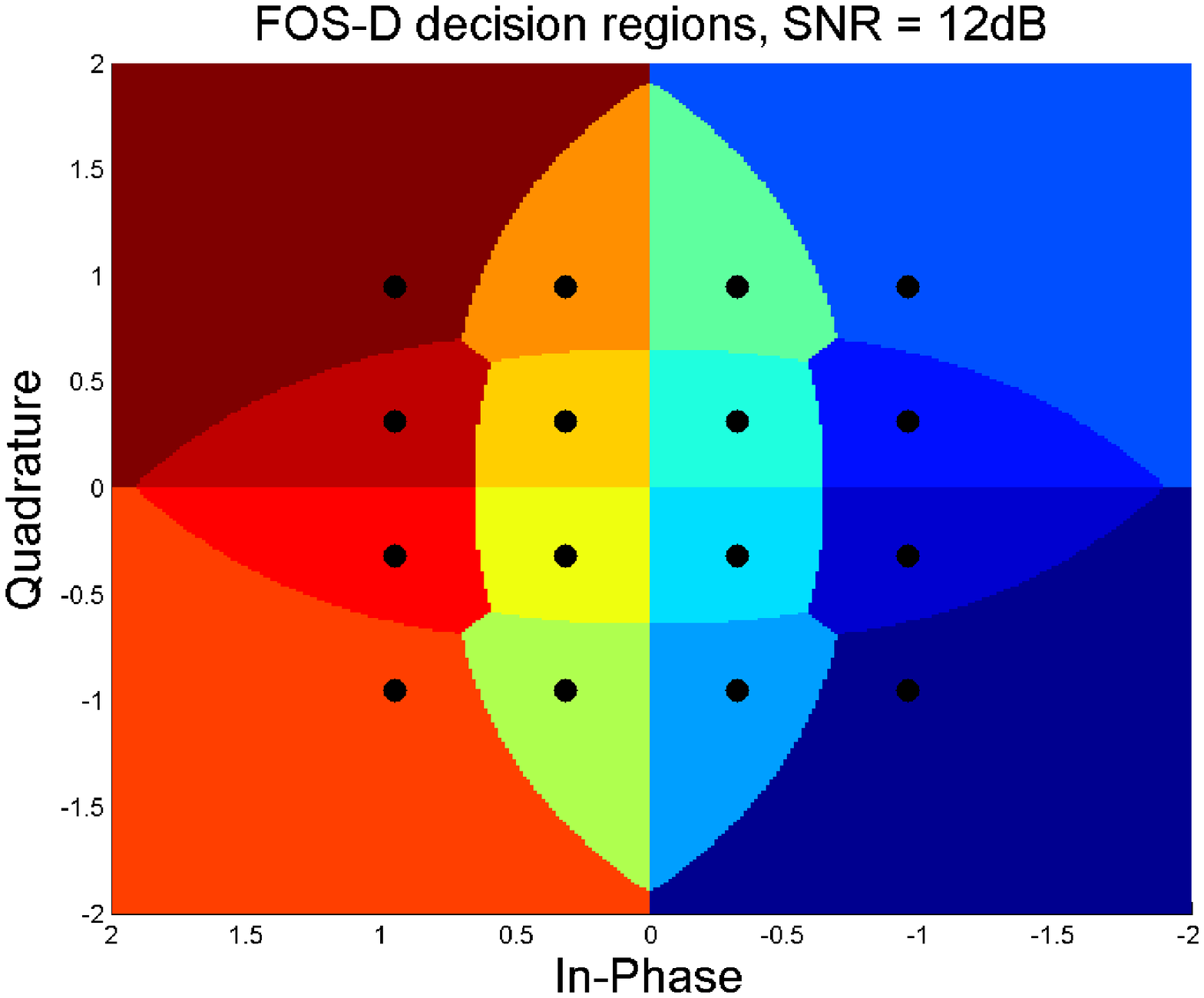} \label{fig:ML_12}
&
\includegraphics[width = 2.1in, keepaspectratio=true]{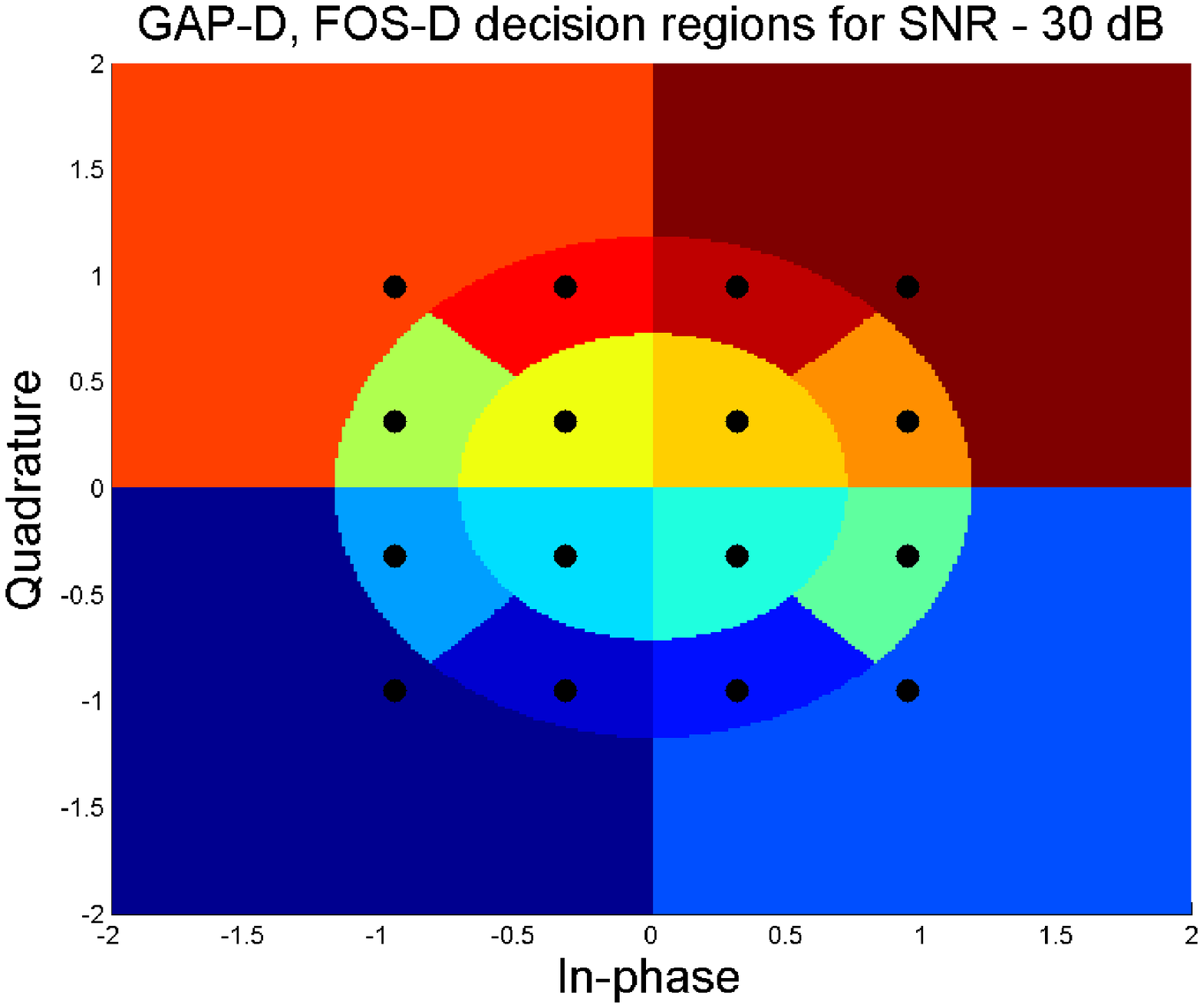} \label{fig:ML_30}
\\
(a) & (b) & (c)\\
\end{tabular}
\caption{Decision regions using 16 QAM, $\sigma_{\text{p}}^{2}=1\times 10^{-2}$ rad$^2$ for (a) GAP-D at SNR = 12 dB (b) FOS-D at SNR = 12 dB  (c) Both GAP-D and FOS-D at SNR = 30 dB }\label{fig:ML_regions}
\end{center}
\end{figure*}

In the following, we derive an ML detector based on a Gaussian assumption for the phase error PDF. Let $p(\theta_{k}|\overline{\textbf{r}}_{k},m_{k} = s_{i})$ denote the PDF of the phase error obtained after compensating the received signal with the estimated phase $\hat{\phi}_{k}$ at time instant $k$. Using a Gaussian PDF assumption for phase error, we have $p(\theta_{k}|\overline{\textbf{r}}_{k},m_{k} = s_{i}) \approx \mathcal{N}(0,\sigma_{\text{p}}^{2} (\overline{\textbf{r}}_{k},m_{k}) )$, i.e., the variance $\sigma_{\text{p}}^{2}$ obtained from the tracker/estimator is a function of $\overline{\textbf{r}}_{k}$ and $m_{k}$, and is known to the detector. 

In order to derive the ML decision based on Gaussian phase error PDF, we rewrite \refeq{eq:mod_ml} as
\begin{eqnarray}\label{eq:mod_ml_new}
L_{i}(k) &=& \int_{-\pi}^{\pi}p(r_{k}|m_{k} = s_{i},\theta_{k})p(\theta_{k}|\overline{\textbf{r}}_{k},m_{k} = s_{i})d\theta_{k} ,\nonumber\\
&\overset{(\text{a})}{=}& \int_{-\pi}^{\pi}p(r_{k}|m_{k} = s_{i},\theta_{k},\overline{\textbf{r}}_{k})p(\theta_{k}|\overline{\textbf{r}}_{k},m_{k} = s_{i})d\theta_{k} ,\nonumber\\
&=& \int_{-\pi}^{\pi}p(r_{k},\theta_{k}|m_{k} = s_{i},\overline{\textbf{r}}_{k})d\theta_{k} ,\nonumber\\
&=& p(r_{k}|m_{k} = s_{i},\overline{\textbf{r}}_{k}),\nonumber\\
&=& p(|r_{k}|, \arg\{r_{k}\}||m_{k}|, \arg\{m_{k}\},\overline{\textbf{r}}_{k}),
\end{eqnarray}
where in step (a), the conditioning with $\overline{\textbf{r}}_{k}$ is valid because $r_{k}$ is independent of $\overline{\textbf{r}}_{k}$, given $m_{k}$ and $\theta_{k}$ (see \refeq{eq:sig_mod_new}). We now study the amplitude and phase of the received signal $r_{k}$ separately. The magnitude of the received signal can be written as follows
\begin{subequations}
\begin{eqnarray}\label{eq:amp_channel_a}
\left| r_{k} \right| &=& \left| m_{k}e^{j\theta_{k}} + n_{k}  \right| \\
&=& \left| \left|m_{k}\right| e^{j\left(\theta_{k} + \text{arg}\{m_{k}\}\right)} + n_{k}^{\prime}e^{j\left(\theta_{k} + \text{arg}\{m_{k}\}\right)}  \right| \\
\label{eq:amp_channel_b}
&=& \left| \left|m_{k}\right| + n_{k}^{\prime}  \right|,  \\
\label{eq:amp_channel_c}
&=& \sqrt{ \left(\left|m_{k}\right| + \text{Re}\{n_{k}^{\prime}\} \right)^{2} + \left(\text{Im}\{n_{k}^{\prime}\} \right)^{2}  } , \\
\label{eq:amp_channel_d}
&\approx&  \left(\left|m_{k}\right| + \text{Re}\{n_{k}^{\prime}\} \right)   ,
\end{eqnarray}
\end{subequations}
where \refeq{eq:amp_channel_b} results from $|e^{j\theta_{k}}| = 1$,  $ n_{k}^{\prime} \triangleq  n_{k}e^{-j\left(\theta_{k} + \text{arg}\{m_{k}\}\right)}$, and the simplification in \refeq{eq:amp_channel_d} arises from the high SNR approximation. The phase of the received signal can be written as
\begin{eqnarray}
\text{arg} \{ r_{k} \} &=& \text{arg} \{ m_{k}e^{j\theta_{k}} + n_{k}  \} , \nonumber\\
&=& \text{arg} \{ \left|m_{k}\right| e^{j\left(\theta_{k} + \text{arg}\{m_{k}\}\right)} + n_{k}^{\prime}e^{j\left(\theta_{k} + \text{arg}\{m_{k}\}\right)} \} , \nonumber\\
&=& \text{arg} \{ \left|m_{k}\right|  + n_{k}^{\prime} \} + \theta_{k} + \text{arg}\{m_{k}\}, \nonumber\\
&=& \arctan{ \frac{\text{Im}  \{n_{k}^{\prime}\} } { \left| m_{k} \right|  + \text{Re}\{n_{k}^{\prime}\}}  } + \theta_{k} + \text{arg}\{m_{k}\} , \nonumber\\
\label{eq:phase_channel_a}
&\approx& { \frac{\text{Im}  \{n_{k}^{\prime}\} } { \left| m_{k} \right|}  } + \theta_{k} + \text{arg}\{m_{k}\} ,
\end{eqnarray}
where, in \refeq{eq:phase_channel_a}, a high SNR approximation has been used.
Now the received signal can be expressed approximately in terms of amplitude and phase r.v.s as follows
\begin{eqnarray}\label{eq:amp_phase_channel}
\text{arg} \{ r_{k} \}
&\approx& \text{arg}\{m_{k}\} +  \frac{\text{Im}  \{n_{k}^{\prime}\} } { \left| m_{k} \right|}   + \theta_{k}  ,\nonumber\\
\left| r_{k} \right|
&\approx& \left|m_{k}\right| + \text{Re}\{n_{k}^{\prime}\}    ,
\end{eqnarray}
where $\text{Im}  \{n_{k}^{\prime}\}  / \left| m_{k} \right|   + \theta_{k}$ is the additive Gaussian noise in the phase channel, and $ \text{Re}\{n_{k}^{\prime}\}$ is the additive noise in the amplitude channel. Thus the observation at the $k$th time instant is equivalently a $2$-tuple $[\left| r_{k} \right| - \left|m_{k}\right|, \text{arg} \{ r_{k} \} -  \text{arg} \{ m_{k} \}  ]$ comprising of amplitude and phase of the received signal. The PDF of this tuple, given $m_{k}$ and  $\overline{\textbf{r}}_{k}$, is a bivariate Gaussian distribution. Conditioning with respect to $\overline{\textbf{r}}_{k}$ occurs because $\sigma_{\text{p}}^{2}$ is the variance (from tracker/estimator) that is a function of $\overline{\textbf{r}}_{k}$. This conditional PDF has mean
\begin{eqnarray*}
\mathbb{E}[\left| r_{k} \right| - \left|m_{k}\right|, \;   \text{arg} \{ r_{k} \} -  \text{arg} \{ m_{k} \} | m_{k},\overline{\textbf{r}}_{k} ] \nonumber\\
= \mathbb{E}[ \text{Re} \{n_{k}^{\prime}\}, \;  \frac{\text{Im}  \{n_{k}^{\prime}\} } { \left| m_{k} \right|}   + \theta_{k}  ] = [0\; 0]^{T},\nonumber
\end{eqnarray*}
and covariance
\begin{align*} \label{eq:var_ml}
\mathbb{E}\left[
\begin{array}{cc}
 |\Re \{n_{k}^{\prime}\}|^{2} &  \Re \{n_{k}^{\prime}\}\left(|m_{k}|\theta_{k} + \Im \{n_{ik}^{\prime} \}\right) \\
\Re \{n_{k}^{\prime}\}\left(\frac{\text{Im}  \{n_{k}^{\prime}\} } { \left| m_{k} \right|}   + \theta_{k} \right) & | \frac{\text{Im}  \{n_{k}^{\prime}\} } { \left| m_{k} \right|}   + \theta_{k} |^{2}
\end{array}
| m_{k},\overline{\textbf{r}}_{k} \right] \nonumber\\
 = \left[
\begin{array}{cc}
N_{0}/2 & 0 \\
0 & \frac{N_{0}}{2\left|m_{k}\right|^{2}} + \sigma_{\text{p}}^{2}
\end{array}
\right].
\end{align*}

Hence, the conditional PDF is written as
\begin{align}
 p(\left| r_{k} \right|, \text{arg} \{ r_{k} \} |m_{k},\overline{\textbf{r}}_{k}) = \frac{e^{\frac{-1}{2}\left( \frac{\left(\left| r_{k} \right| - \left|m_{k}\right|\right)^{2} }{\frac{N_{0}}{2}}   + \frac{\left(\text{arg} \{ r_{k} \} - \text{arg}\{m_{k}\}\right)^{2} }{\frac{N_{0}}{2\left|m_{k}\right|^{2}} + \sigma_{\text{p}}^{2}}\right)}}{2\pi \sqrt{\left(\frac{N_{0}}{2}\right)\left(\frac{N_{0}}{2\left|m_{k}\right|^{2}} + \sigma_{\text{p}}^{2} \right) } },\nonumber\\
\end{align}
where $\sigma_{p}^{2}$ is a function of $m_{k}$ and $\overline{\textbf{r}}_{k}$, and is known to the detector. Based on the joint observation of phase and amplitude of the received signal,  the ML decision rule can be expressed using \refeq{eq:mod_ml_new} as
\begin{eqnarray}\label{eq:ml_gaussian}
\hat{s}_{k} &=&  \underset{i\in \{1,\ldots,C\}}{\operatorname{argmax}}\, P(\left| r_{k} \right|, \text{arg} \{ r_{k} \} |s_{i},\overline{\textbf{r}}_{k}) \nonumber\\
&=&  \underset{i\in \{1,\ldots,C\}}{\operatorname{argmin}}\, \frac{\left(\left|r_{k}\right| - \left|s_{i}\right|\right)^{2}}{N_{0}/2}  \\
&+&  \frac{\left(\arg\{r_{k}\} - \arg\{s_{i}\}\right)^{2}}{\sigma_{\text{p}}^{2} + \frac{N_{0}}{ 2{\left|s_{i}\right|}^{2}}} + \log\left(\sigma_{\text{p}}^{2} + \frac{N_{0}/2}{{ \left|s_{i}\right|^{2}}}\right),\nonumber
\end{eqnarray}
This ML decision rule can be regarded as a joint amplitude and phase detector that involves a weighted combination of the amplitude and phase differences between the received signal and all symbols in the constellation. The $\log(\cdot)$ term in \refeq{eq:ml_gaussian} may be removed since it has negligible influence on the decision in practical scenarios..

We denote the detector in \refeq{eq:ml_gaussian} as \emph{GAP-D}. Note that by splitting the received signal $r_{k}$ into its amplitude and phase components, the derivation of the ML does not involve an explicit marginalization with respect to the phase error. In order to visually understand the decision rule in \refeq{eq:ml_gaussian}, the decision regions employing GAP-D and FOS-D are presented in Fig. \ref{fig:ML_regions}(a), \ref{fig:ML_regions}(b) and \ref{fig:ML_regions}(c) for 16-QAM, a fixed phase error variance of $\sigma_{\text{p}}^{2}=1\times10^{-2}$ rad$^2$, and SNR values of 12 dB and 30 dB. Some properties of GAP-D are as follows:
\begin{itemize}
\item It can be verified that GAP-D reduces to the Euclidean distance based detector in its polar form when $\sigma_{\text{p}}^{2} \rightarrow 0$ in \refeq{eq:ml_gaussian}.
\item As $\sigma_{p}^{2} \rightarrow \infty$ for a given SNR, signal points with equal energy become indistinguishable, and the detector is essentially an amplitude (or energy) detector. This corresponds to the case of completely incoherent systems.
\item In high SNR scenarios $N_{0} \rightarrow 0$, GAP-D discards symbols that have large amplitude difference relative to the received signal amplitude. It only considers a smaller subset of symbols that have relatively small amplitude difference with respect to the received signal. The transmitted symbol is determined from this subset based on phase difference with respect to the received signal.
\item The ML detector remains unchanged if both the signal and noise are scaled by the same constant value for a given phase error variance. In summary, the ratio between the AWGN variance and phase noise $\sigma_{\text{p}}^{2}$ characterizes the ML decision rule for such systems.
\end{itemize}

\subsection{Two-Step Amplitude-Phase Detector (TS-D)}

We now consider the amplitude-phase detector that was derived in \cite{kahn07} where in signal detection is carried out in two step. This detector maybe viewed as a sub-optimal version of GAP-D. The first step involves an amplitude detection based on the received signal. This renders a subset $\mathcal{C}_{\text{amp}}$ of the symbol constellation $\mathcal{C}$ that are closest to the received signal in terms of its amplitude. Then based on $\mathcal{C}_{\text{amp}}$, phase detection is performed wherein the final symbol chosen is closest to the received signal in terms of phase. As we shall see in the sequel, this detector has SEP performance similar to that of GAP-D at high SNR, where amplitude ambiguity of the received signal is small and the uncertainty primarily lies in phase. This detector does not make any explicit assumptions on the phase error PDF. We denote this detector as \emph{TS-D}.


\subsection{Metric as Weighted Sum of Central Moments}

For the ML detector presented in \refeq{eq:ml_fosc} and \refeq{eq:ml_gaussian}, assumptions are made about the PDF of the phase error obtained after compensation of the received signal with a phase estimate from a phase-tracker/smoother. Note that the phase error, after compensation by a filter, is approximately Gaussian only in some scenarios. This includes cases where the amplitudes of the transmitted symbols are the same, when the phase estimation is conditioned on the amplitude of the transmitted symbols or when the estimator is locked, i.e, the estimated phase closely follows the actual phase. In fact, an accurate derivation of the phase error PDF for any constellation is an analytically difficult problem. We present a brief argument for why this is the case in general. Let $P(\theta_{k})$ denote the PDF of the phase error $\theta_{k}$, which is obtained after  compensating using a phase-tracker/smoother in the $k$th time instant for a general non-equal energy constellation $\mathcal{C}$. Then we have
\begin{eqnarray} \label{eq:sum_PDF}
P(\theta_{k}) =  \sum_{\overline{\textbf{r}}_{k}} \sum_{s_{i} \in \mathcal{C} } P(\theta_{k}|s_{i},\overline{\textbf{r}}_{k})P(\overline{\textbf{r}}_{k}|s_{i})P(s_{i}).
\end{eqnarray}
Upon using any practical trackers/smoothers, the posteriori PDFs $P(\theta_{k}|s_{i},\overline{\textbf{r}}_{k})$ \refeq{eq:sum_PDF} will have statistics that depend on the instantaneous SNR or the magnitude of $s_{i}$ and $\overline{\textbf{r}}_{k}$. Hence, the PDF in \refeq{eq:sum_PDF} becomes a mixture of posteriori PDFs, which in general may not be analytically tractable. Therefore, it is of interest to determine an (approximate) ML decision rule without making restrictive assumptions on the distribution of the phase error $\theta_{k}$.

Adopting the system model in \refeq{eq:sig_mod_new}, consider the problem of data detection in the $k$th time slot after compensating the received signal with a phase estimate \refeq{eq:mod_ml}. Assume $\theta_{k}$ to be the phase error that is drawn from an arbitrary probability distribution. Consider the likelihood function $f(\theta_{k})\triangleq p(r_{k}|m_{k} = s_{i},\theta_{k})$ given as
\begin{equation*}\label{eq:f_theta}
f(\theta_{k})= \frac{e^{-\frac{|r_{k} - s_{i}e^{j\theta_{k}}|^2}{2N_{0}}}}{(2\pi N_{0})^{1/2}},
\end{equation*}
Then, by performing Taylor series expansion of the likelihood function $f(\theta_{k})$ about $\theta_{k} = \hat{\theta}_{k}$ ( where $\hat{\theta}_{k}$ the mean of $\theta_{k}$), \refeq{eq:mod_ml} can be rewritten as
\begin{align}\label{eq:sum_moments}
\underset{i\in \{1,\ldots,C\}}{\operatorname{argmax}} L_{i}(k) =& \underset{i\in \{1,\ldots,C\}}{\operatorname{argmax}}  \int_{-\pi}^{\pi}
\left[\frac{f(\hat{\theta}_{k})}{0!}+ \frac{f^{\{1\}}(\hat{\theta}_{k})}{1!}\right.\nonumber\\
& \left. \times (\theta_{k} - \hat{\theta}_{k}) + \frac{f^{\{2\}}(\hat{\theta}_{k})}{2!}(\theta_{k} - \hat{\theta}_{k})^{2} + \ldots \right]\nonumber\\
& \times {p(\theta_{k}|m_{k } = s_{i}, \overline{\textbf{r}}_{k})}d\theta_{k},\nonumber\\
=& \underset{i\in \{1,\ldots,C\}}{\operatorname{argmax}} \frac{1}{(2\pi N_{0})^{1/2}}\left[\frac{f(\hat{\theta}_{k})M_{0}}{0!} \right. \\& + \left.\frac{f^{\{1\}}(\hat{\theta}_{k})M_{1}}{1!} \ldots + \frac{f^{\{n\}}(\hat{\theta}_{k})M_{n}}{n!} + \ldots  \right]. \nonumber
\end{align}
That is, the decision rule in \refeq{eq:mod_ml} is equivalent to the maximization of the weighted sum of $M_{j}, j \in \mathbb{Z}^{+}$ over $s_{i} \in \{1,\ldots,C\}$. Here, $M_{j}$ is the $j$th central moment of the phase error PDF, $p(\theta_{k}|\overline{\textbf{r}}_{k},m_{k} = s_{i})$, and
$f^{\{n\}}(\hat{\theta}_{k})$ is the $n$th derivative of $f(\theta_{k})$ evaluated about $\theta_{k} = \hat{\theta}_{k}$. For the Taylor series expansion in \refeq{eq:sum_moments} to converge to $f(\theta_{k})$ for all values of $\theta_{k}$, it is required that $f(\theta_{k})$ be an analytic function in $\theta_{k}$, $\forall\theta_{k} \in \mathbb{R}$. This can be easily shown as was done in our prior work \cite{raj11}. For brevity, we refrain from reproducing the proof of analyticity here.

In deriving \refeq{eq:sum_moments}, no assumptions are made on the distribution/statistics of $\theta_{k}$ or SNR. Thus, the problem of determining the ML detector is reduced to estimating the central moments of the posteriori PDF of $\theta_{k}$. In its exact form, the new detector incurs high computational complexity on the receiver as it requires knowledge of all central moments of the posteriori PDF. Also computing the $n$th derivative of $f(\hat{\theta}_{k})$  is complex when $n$ is large.  As we shall see in the sequel, the parametric form of the ML detection rule in \refeq{eq:sum_moments} allows a simple approximation by using a finite number of terms that achieves good SEP performance relative to the ML decision rule. However, assumptions have to be made about the SNR and the phase noise variance when the ML detector \refeq{eq:sum_moments} is truncated, which is investigated in the sequel.

\setcounter{equation}{21}
\begin{figure*}[!hb]
\begin{eqnarray}
    \label{eq:mean_var_eta}
a_{ij} \triangleq \frac{\left(\left|s_{i}\right| - \left|s_{j}\right|\right)^{2}}{N_{0}/2}, \;
b_{ij} &\triangleq& \frac{\left(\arg\{s_{i}\} - \arg\{s_{j}\}\right)^{2}}{\sigma_{\text{p}}^{2} + \frac{N_{0}}{ 2{\left|s_{j}\right|}^{2}}},\;
c_{ij} \triangleq \frac{\sigma_{\text{p}}^{2} + \frac{N_{0}}{ 2{\left|s_{i}\right|}^{2}}}{\sigma_{\text{p}}^{2} + \frac{N_{0}}{ 2{\left|s_{j}\right|}^{2}}},\;
y_{ij} \triangleq  \log{\frac{\left({ \left|s_{i}\right|^{2}}\sigma_{\text{p}}^{2} + \frac{N_{0}}{2}\right)}{\left({ \left|s_{j}\right|^{2}}\sigma_{\text{p}}^{2} + \frac{N_{0}}{2}\right)}}\nonumber\\
\mathbb{E}\{\eta_{ij}|s_{i}\} &=&  1 - (a_{ij} + b_{ij} + c_{ij}),\nonumber\\
\mathbb{E}\left\{\left(\eta_{ij} - \mathbb{E}\{\eta_{ij}|s_{i}\}\right)^{2}|s_{i}\right\} &=& \mathbb{E}\{ L_{i}^{2}|s_{i}\} +  \mathbb{E}\{ L_{j}^{2} |s_{i}\} - 2\mathbb{E}\{ L_{i}L_{j} |s_{i}\} - \left(  \mathbb{E}\{\eta_{ij}|s_{i}\} \right)^{2},\nonumber\\
  &=& 2  + 4a_{ij} + 2c_{ij}^{2}  + 4b_{ij}c_{ij} - 4c_{ij}.
  \end{eqnarray}
\end{figure*}

\subsubsection{Truncation of the Sum-of-Central-Moments Metric (SOM-D)}

A straightforward approach to approximating the ML decision rule is to truncate the number of terms in the Taylor series expansion, $n$, and investigate the various scenarios where the approximate decision rule achieves SEP performance close to that of the ML decision rule. Consider the case where the Taylor series is truncated to $n=2$; i.e., only the first three terms in the optimal decision rule in \refeq{eq:sum_moments} are considered. This case is particularly interesting as it corresponds to scenarios where the posteriori distribution of $\theta_{k}$ is unknown except for its mean and variance. We first define an approximate SBS detection rule for data over AWGN channel for $n = 2$ as
\setcounter{equation}{17}
\begin{eqnarray}\label{eq:sum_moments_2}
\underset{i\in \{1,\ldots,C\}}{\operatorname{argmax}}\, A_{i,2}(k)  =&  \underset{i\in \{1,\ldots,C\}} {\operatorname{argmax}}\, \left[\frac{f(\hat{\theta}_{k})M_{0}}{0!} \right. \left. + \frac{f^{\{2\}}(\hat{\theta}_{k})\sigma_{\text{p}}^{2}}{2!}\right],
\end{eqnarray}
where $\frac{f^{\{1\}}(\hat{\theta}_{k})}{1!}M_{1} = 0$ since the first central moment of any r.v. is zero. This detector is referred to as \emph{SOM-D}. The second-order approximate ML rule in \refeq{eq:sum_moments_2} consists of two terms; the first term is the zeroth order term from the Taylor series and is identical to the Euclidean distance based symbol detection rule. The second term is the variance of the posteriori PDF of $\theta_{k}$ weighted by the second derivative of $f(\theta_{k}) = p(r_{k}|m_{k} = s_{i},\theta_{k})$, which intuitively gives a measure of its sharpness or curvature about $r_{k}=s_{i}e^{j\hat{\theta}_{k}}$. 

In our prior work \cite{raj11}, an upper bound on the error for this approximation was derived using the Taylor remainder formula. Here we only recall the scenarios where the approximate decision rule would have a large error relative to the original ML decision rule \refeq{eq:sum_moments}: (a) the error in approximation is inversely proportional to AWGN channel noise variance, (b) the error is directly proportional to the variance of posteriori PDF of $\theta_{k}$. Hence SOM-D with $2$ moments as in \refeq{eq:sum_moments_2} is interesting for the following scenarios: (a) low-to-medium SNR, low-to-high phase noise, and (b) low-to-medium phase noise, low-to-high SNR.

Finally, we remark that the SOM-D may be truncated to $n$ terms at the cost of computational complexity. It is also possible to determine the minimum number of moments $n$ required to make the approximate SOM-D close to the original rule in \refeq{eq:sum_moments}. This value can be arbitrarily large for high phase noise variance and SNR. Due to space restrictions, we refrain from reproducing those results here, and refer interested readers to \cite{raj11} for a deeper analysis of this detector.

\section{SEP Analysis for GAP-D}
\label{sec:sep_ana}

In this section, we derive analytical expressions for the SEP performance of GAP-D for arbitrary constellations. Consider a constellation with $M$ symbol points that are equally likely and a received signal model as in \refeq{eq:sig_mod_new}, where $\theta_{k}$ denotes the phase error as before. The symbol error probability $P_{e}$ for this constellation is upper-bounded by averaging over all pair-wise symbol error probabilities (union bound) \cite{proakis} as follows
\setcounter{equation}{18}
\begin{eqnarray}\label{eq:error_bound}
P_{e} \leq \frac{1}{M} \sum_{i=1}^{M} \sum_{j = 1, j \neq i }^{M} P\left( E_{ij} \right),
\end{eqnarray}
where $P\left( E_{ij} \right)$ is the probability of a pair-wise symbol error event. This corresponds to an the event where the received symbol is not detected as symbol $i, \, i \in \{1,\ldots,M\}$, given symbol $i$ has been transmitted. More precisely, based on the ML detector derived, the probability of pair-wise symbol error event can be expressed as follows (dropping time index $k$)
\begin{eqnarray}
P\left( E_{ij} \right) &=& Pr \left( L_{j} < L_{i} | s_{i} \right),\nonumber
\end{eqnarray}
where the pair-wise symbol error event corresponds to metric $L_{j}$ associated with symbol $j$ being lower than that for symbol $i$. Let us examine the event corresponding to $\left(L_{i} - L_{j} > 0 | s_{i}\right)$ by first evaluating the difference $\eta_{ij} = L_{i} - L_{j}$ that, using \refeq{eq:ml_gaussian} is given as
\begin{eqnarray}
 \eta_{ij} &=& \log{\frac{\left(\sigma_{\text{p}}^{2} + \frac{N_{0}}{2{ \left|s_{i}\right|^{2}}}\right)}{\left(\sigma_{\text{p}}^{2} + \frac{N_{0}}{2{ \left|s_{j}\right|^{2}}}\right)}} +  \frac{\left(\left|r\right| - \left|s_{i}\right|\right)^{2}}{N_{0}/2} - \frac{\left(\left|r\right| - \left|s_{j}\right|\right)^{2}}{N_{0}/2}   \nonumber\\
&+& \frac{\left(\arg\{r\} - \arg\{s_{i}\}\right)^{2}}{\sigma_{\text{p}}^{2} + \frac{N_{0}}{ 2{\left|s_{i}\right|}^{2}}} - \frac{\left(\arg\{r\} - \arg\{s_{j}\}\right)^{2}}{\sigma_{\text{p}}^{2} + \frac{N_{0}}{ 2{\left|s_{j}\right|}^{2}}}.\nonumber
\end{eqnarray}
Given symbol $s_{i}$ has been transmitted, the amplitude and phase of the received signal $r$ from \refeq{eq:amp_phase_channel} are as follows
\begin{eqnarray}\label{eq:rxd_model}
\left| r \right|
&\approx&  \left|s_{i}\right| + \text{Re}\{n^{'}\} ,\nonumber\\
\text{arg} \{ r \}
&\approx& { \frac{\text{Im}  \{n^{'}\} } { \left| s_{i} \right|}  } + \theta + \text{arg}\{s_{i}\}.
\end{eqnarray}
Based on \refeq{eq:rxd_model}, given the transmitted symbol $s_{i}$, $\eta_{ij}$  simplifies as follows
\begin{align}\label{eq:eta_ij}
 \eta_{ij} &=& \log{\frac{\left(\sigma_{\text{p}}^{2} + \frac{N_{0}}{2{ \left|s_{i}\right|^{2}}}\right)}{\left(\sigma_{\text{p}}^{2} + \frac{N_{0}}{2{ \left|s_{j}\right|^{2}}}\right)}} - \underbrace{ \frac{2(\left|s_{i}\right|-\left|s_{j}\right|)\text{Re}\{n^{'}\} + \left(\left|s_{i}\right| - \left|s_{j}\right| \right)^{2}}{N_{0}/2} }_{\triangleq {V_{1}}} \nonumber\\  &+& \underbrace{  \frac{\left(   { \frac{\text{Im}  \{n^{'}\} } { \left| s_{i} \right|}  } + \theta    \right)^{2}}{\sigma_{\text{p}}^{2} + \frac{N_{0}}{ 2{\left|s_{i}\right|}^{2}}} }_{\triangleq {V_{2}}} - \underbrace{ \frac{\left({ \frac{\text{Im}  \{n^{'}\} } { \left| s_{i} \right|}  } + \theta + \arg\{s_{i}\} - \arg\{s_{j}\}\right)^{2}}{\sigma_{\text{p}}^{2} + \frac{N_{0}}{ 2{\left|s_{j}\right|}^{2}}} }_{\triangleq {V_{3}}}.\nonumber\\
\end{align}

In \refeq{eq:eta_ij}, given that symbol $s_{i}$ has been transmitted, the term denoted as $ V_{1}$ is a Gaussian r.v.. Consider the term denoted as $ V_{2} - V_{3}$ that can simplified as
\begin{eqnarray}
V_{2}-V_{3} &=&   \frac{\left(   { \frac{\text{Im}  \{n^{'}\} } { \left| s_{i} \right|}  } + \theta    \right)^{2}}{\sigma_{\text{p}}^{2} + \frac{N_{0}}{ 2{\left|s_{i}\right|}^{2}}}  -  \frac{\left({ \frac{\text{Im}  \{n^{'}\} } { \left| s_{i} \right|}  } + \theta \right)^{2}}{\sigma_{\text{p}}^{2} + \frac{N_{0}}{ 2{\left|s_{j}\right|}^{2}}}  \nonumber\\ &-& 2 \frac{{\left({ \frac{\text{Im}  \{n^{'}\} } { \left| s_{i} \right|}  } + \theta \right)}\left(\arg\{s_{i}\} - \arg\{s_{j}\}\right)}{\sigma_{\text{p}}^{2} + \frac{N_{0}}{ 2{\left|s_{j}\right|}^{2}}}  \nonumber\\ &-& \frac{\left(\arg\{s_{i}\} - \arg\{s_{j}\}\right)^{2}}{\sigma_{\text{p}}^{2} + \frac{N_{0}}{ 2{\left|s_{j}\right|}^{2}}}.\nonumber
\end{eqnarray}
At high SNR, $\sigma_{\text{p}}^{2} + {N_{0}}/{ 2{\left|s_{i}\right|}^{2}}$ $\approx$ $\sigma_{\text{p}}^{2} + {N_{0}}/{ 2{\left|s_{j}\right|}^{2}}$, and hence the term consisting ${\text{Im}  \{n^{'}\} } /{ \left| s_{i} \right|}   + \theta$ dominates. As a consequence, $V_{2}-V_{3} $ can be approximated as a Gaussian r.v.. Hence $\eta_{ij}$ is approximated as a Gaussian r.v. whose mean and covariances are derived in \refeq{eq:mean_var_eta}. As we shall see in the simulation section, this approximation is appropriate for a wide range of SNR values. Hence, the probability of the error event $\text{Pr}\left(L_{i} - L_{j} > 0 | s_{i}\right) = $ $\text{Pr}\left(\eta_{ij} > 0 | s_{i}\right)$ can be approximated in high SNR scenarios as
\setcounter{equation}{22}
\begin{eqnarray}\label{eq:p_error_etaij}
\text{Pr}\left(\eta_{ij} > 0 | s_{i}\right) \approx \mathcal{Q}  \left( \frac{y_{ij} - \mathbb{E}\{\eta_{ij}|s_{i}\} }{\sqrt{  \mathbb{E}\left\{\left(\eta_{ij} - \mathbb{E}\{\eta_{ij}|s_{i}\}\right)^{2}|s_{i}\right\}  }} \right).
\end{eqnarray}
Applying the union bound in \refeq{eq:error_bound}, the probability of error for this detector is upper-bounded as
\begin{eqnarray}\label{eq:p_error}
P_{e} \leq \frac{1}{M}\sum_{i = 1}^{M}\sum_{j = 1,j \neq i}^{M}  \mathcal{Q}  \left( \frac{y_{ij} - \mathbb{E}\{\eta_{ij}|s_{i}\} }{\sqrt{  \mathbb{E}\left\{\left(\eta_{ij} - \mathbb{E}\{\eta_{ij}|s_{i}\}\right)^{2}|s_{i}\right\}  }} \right).
\end{eqnarray}

\begin{figure*}[!ht]
\begin{center}
\begin{tabular}{cc}
\includegraphics[width = 3.0in, keepaspectratio=true]{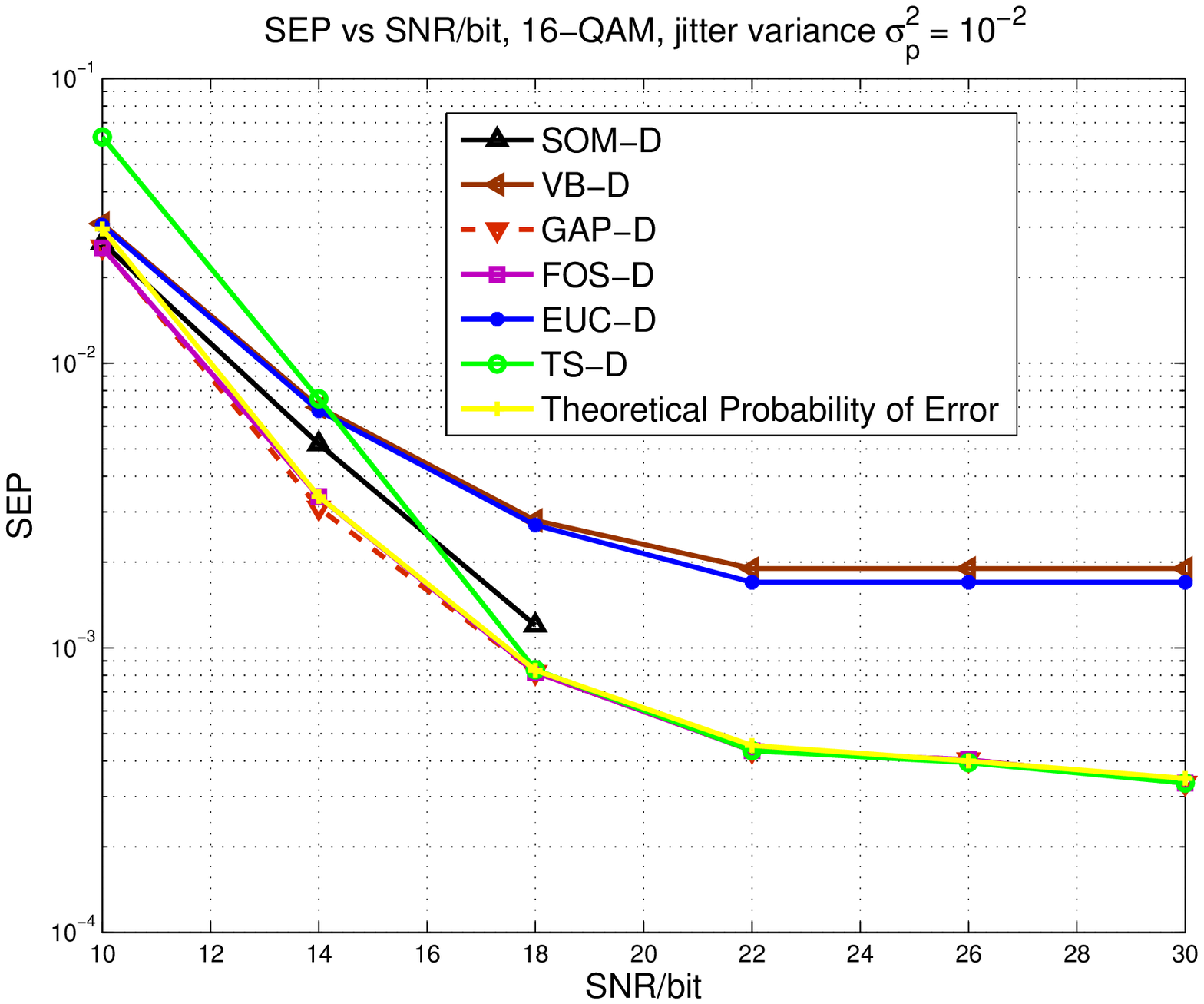}
&
\includegraphics[width = 3.0in, keepaspectratio=true]{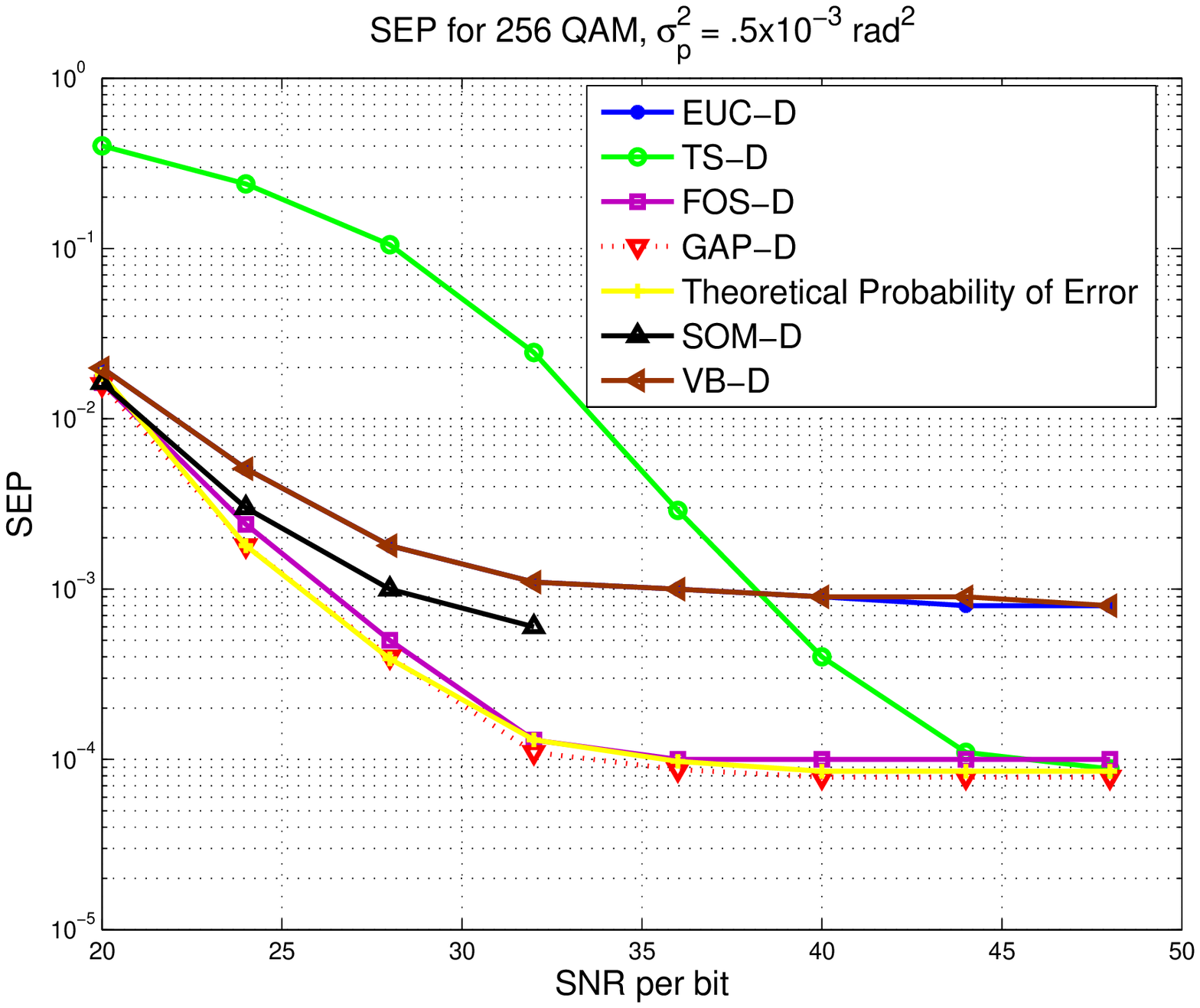}
\\
(a) & (b)\\
\end{tabular}
\caption{Comparison of SEP performance between the various detectors and theoretical SEP for different constellations and $\sigma_{\text{p}}^{2}$ values - (a) 16 QAM and $\sigma_{\text{p}}^{2}=10^{-2} \text{rad}^{2}$ (b) 256 QAM and  $\sigma_{\text{p}}^{2}=0.5 \times 10^{-3} \text{rad}^{2}$ }\label{fig:metric_comp}
\end{center}
\end{figure*}

The argument of the $\mathcal{Q}$ function in \refeq{eq:p_error} can be viewed as a normalized distance measure (analogous to Euclidean distance in AWGN), which can be used to determine the nearest neighbors of a given symbol in the presence of phase noise.

\subsection{SEP at High SNR and Error Floors }
In order to determine SEP at high SNR and the error floor for a given constellation, the probability of error \refeq{eq:p_error} can be simplified in high SNR scenarios by evaluating $\underset{N_{0} \to 0} \lim \frac{\left(\left|s_{i}\right| - \left|s_{j}\right|\right)^{2}}{N_{0}/2} \triangleq $ $\underset{N_{0} \to 0} \lim a_{ij} = 0$  for equal-energy symbol points and letting $\underset{N_{0} \to 0}\lim a_{ij} \to \infty$ for non-equal energy points. Hence, the value of $Pr\left(\eta_{ij} > 0 | s_{i}\right)$ in \refeq{eq:p_error_etaij} tends to zero for asymptotically high SNR for any pair of symbol points with non-equal energy since $ \mathcal{Q}(\cdot)$ is a decreasing function of its argument. However, for any pair of symbol points with equal energy, $\text{Pr}\left(\eta_{ij} > 0 | s_{i}\right)$ from \refeq{eq:p_error_etaij} simplifies in high SNR as follows
\begin{eqnarray}
\underset{N_{0} \to 0}\lim  \text{Pr}\left(\eta_{ij} > 0 | s_{i}\right)  = \mathcal{Q} \left( \sqrt{\left(\frac{\left(\arg\{s_{j}\} - \arg\{s_{i}\}\right)^{2}}{\sigma_{\text{p}}^{2}  }  \right)} \right).
\end{eqnarray}
Hence, applying the union bound in \refeq{eq:error_bound} by considering only those pairs of symbols with equal amplitude (energy), we obtain the error floor as
\begin{align} \label{eq:error_floor}
\underset{N_{0} \to 0}\lim P_{e} \leq \frac{1}{M}\sum_{i = 1}^{M}\sum_{\substack{j = 1,j \neq i,\\ |s_{i}|=|s_{j}|}}^{M} \mathcal{Q} \left( \sqrt{\left(\frac{\left(\arg\{s_{j}\} - \arg\{s_{i}\}\right)^{2}}{\sigma_{\text{p}}^{2}  }  \right)} \right).\nonumber\\
\end{align}
Note that the residual phase noise variance, $\sigma_{p}^{2}$, is generally a function of SNR. Hence as SNR increases, $\sigma_{p}^{2}$ reduces until it reaches a limit value, implying that the floor level depends on the phase estimation algorithm. Additionally, we can make some quick deductions about constellations and error floors from \refeq{eq:error_floor}. It can be inferred that a constellations with points that are all at different distances from the origin (like a spiral shaped constellation) may have a very low error floor. Further, if the points in the constellation have the same amplitude, then they should be separated by a large angular distance in order to avoid an error floor.

\section{Simulations}
\label{sec:results}

In this section, we present performance results of the proposed detectors for uncoded data in the presence of strong phase noise in terms of SEP versus $E_{b}/N_{0}$ (SNR per bit), where $E_{b}$ is energy per bit. Note that the uncoded case has been considered only for illustration purposes, and these metrics can be used for detection in coded system as well \cite{imai04}. We consider two scenarios for evaluating the performance of the proposed detectors -
\begin{itemize}
\item In the first scenario, the phase error PDF is considered to be Gaussian  \cite{fos73}, \cite{imai04}, \cite{kram11}.
\item In the second scenario, the phase noise is modeled as a Wiener process (refer to \refeq{eq:wiener}) \cite{cola05}, \cite{vb09}. At the receiver, an estimator is implemented followed by detection using the metrics presented in section \ref{sec:analysis}.
\end{itemize}

\subsection{Gaussian PDF Phase Error}
In the first scenario, the phase error PDF is considered to be Gaussian i.i.d. \cite{fos73}, \cite{imai04}, \cite{kram11} with variance $\sigma_{\text{p}}^{2}$. Two modulation schemes with relatively different constellation order are considered for study: $16-$QAM and $256-$QAM constellations. For the numerical example presented in Fig. \ref{fig:metric_comp}(a), transmitted symbols are drawn uniformly from a $16-$QAM constellation, and the variance of the phase error is fixed as $\sigma_{\text{p}}^{2}=10^{-2} \text{rad}^{2}$ \cite{imai04}. For the example presented in Fig. \ref{fig:metric_comp}(b), transmitted symbols are drawn from a $256-$QAM constellation, and the variance is fixed as $\sigma_{\text{p}}^{2}=.5 \times 10^{-3} \text{ rad}^{2}$.

Given the high variance of phase error, substantial gains can be realized while using GAP-D and FOS-D compared to EUC-D and VB-D for both constellations \ref{fig:metric_comp}(a) and (b). The performance of FOS-D and GAP-D are almost the same for all SNR and phase noise variance values considered. This is because the Tikhonov PDF and the Gaussian PDF are identical for phase noise variance values of practical interest \cite{vit63}. From simulating the performance of SOM-D (with $2$ moments) for 16-QAM, we observe that its SEP performance is better than VB-D and EUC-D till about $18$ dB, beyond which its performance blows up. Recall that the error of truncation of SOM-D is large in scenarios of high SNR and phase noise variance, and more moments have to be incorporated in the decision rule. For the case of 256-QAM constellation with SOM-D (with $2$ moments), a similar observation is made beyond an SNR of $34$ dB. For both constellations, the performance of the TS-D converges with that of the GAP-D and FOS-D in the high SNR regime. This is expected since as the SNR increases, ambiguity in the amplitude component tends to zero, and the perturbation occurs only in the phase component. The convergence of its performance is faster in the case of $16$-QAM compared to $256$-QAM, given that the former is relatively less dense.

In Fig. \ref{fig:metric_comp}(a) and (b), we plot the theoretical probability of error derived in \refeq{eq:p_error}, which is seen to be a tight upper bound for the SNR values considered. Observe that in contrast to the traditional union bound theory, the bound is also tight for high error rates, which can be explained as follows: the contribution of the nearest neighbors to the union bound, for a given transmitted symbol, is significantly more dominant than the contributions from all the other symbols, when we have both phase noise and AWGN. This is because of the directional nature of phase noise. In general, the bound can be seen to be tight whenever the phase  noise variance dominates the AWGN variance. From these figures, it can also be seen that the error floor observed at high SNR can be accurately predicted using the analytical result in \refeq{eq:error_floor}. The error floor arises in QAM constellations because of the presence of symbol points that are of the same amplitude, and are not separated by large angular distances.

\subsection{Wiener Phase Noise Process}

In the second scenario considered for simulations, the transmitted symbols are assumed to be affected by Wiener phase noise process. This type of noise typically arises from free running oscillators at the transmitter and receiver \cite{cola05}, \cite{vb09} and is modeled in the $k$th time instant as
\begin{eqnarray}\label{eq:wiener}
\phi_{k} = \phi_{k-1} + \Delta_{k},
\end{eqnarray}
where $\phi_{k}$ at $k=0$ is a uniform r.v., and $\Delta_{k}\sim \mathcal{N}(0,\sigma_{\Delta}^{2})$ is the innovation component of the random walk process. The signal received at a time instant is fed to an extended Kalman filter (EKF) \cite{vb09} to track its phase. The choice of EKF to be the phase tracker is non-preferential in that we could use any algorithm like particle filter or PLL for tracking $\phi_{k}$. Note that the EKF has structure and performance similar to that of a PLL and other MAP estimation algorithms \cite{pll1}, and are approximate solvers for the optimal nonlinear phase estimation problem \cite{pll3}.

Phase estimate $\hat{\phi}_{k}$ is first used to rotate the received signal (thereby compensating for the Wiener phase noise). Then, this estimate along with its variance (from the EKF), are used in the detector to compute aposteriori symbol probabilities (denoted as $P_{\text{app}}(s_{k})$) using the metrics: GAP-D, FOS-D, SOM-D, COL-D, VB-D and EUC-D. Note that the variance of the phase estimate  $\sigma_{\text{p},k}^{2}$ in time instant $k$, is a function of $m_{k}$ and $\overline{\textbf{r}}_{k}$, and therefore changes with each time instant. Using aposteriori symbol probabilities, a soft symbol $\hat{s}_{k}$ is computed and fed back to the EKF as follows
\begin{eqnarray}
\hat{s}_{k} =  \underset{\forall s_{k} \in\mathcal{C}} {\sum} s_{k}P_{\text{app}}(s_{k}).\nonumber
\end{eqnarray}
\begin{figure}[!ht]
\begin{center}
\includegraphics[width = 3.5in, keepaspectratio=true]{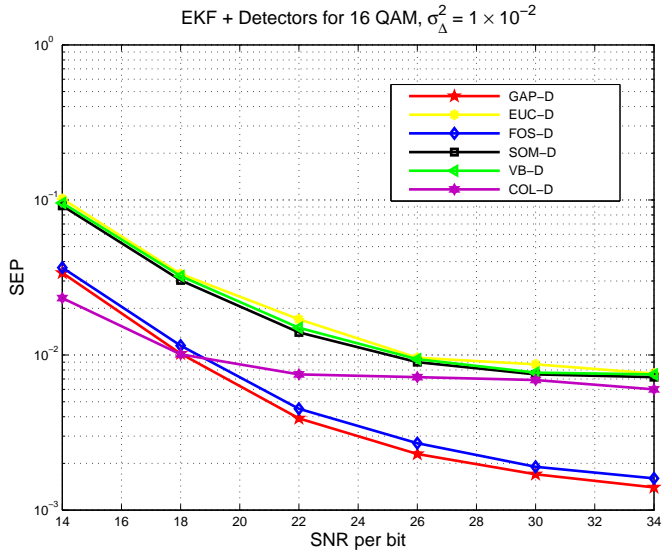}
\caption{Comparison of SEP performance between the various detectors and theoretical SEP for $16$-QAM, $\sigma_{\Delta}^{2}=10^{-2} \text{rad}^{2}$. }\label{fig:ekf_comp_16}
\end{center}
\end{figure}

\begin{figure}[!ht]
\begin{center}
\includegraphics[width = 3.5in, keepaspectratio=true]{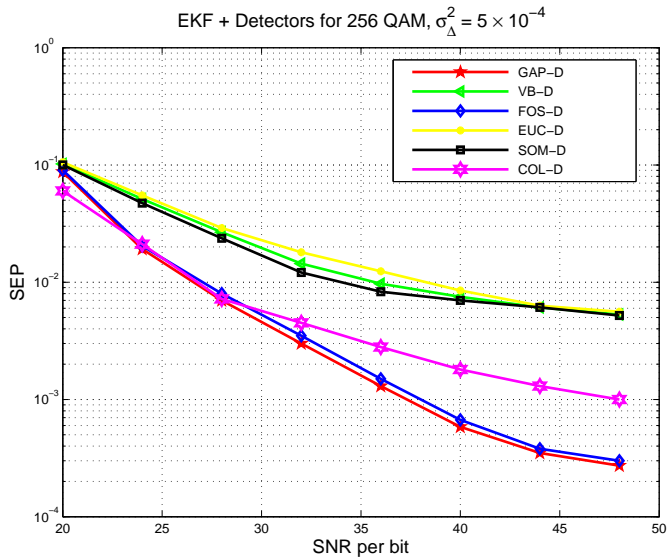}
\caption{Comparison of SEP performance between the various detectors and theoretical SEP for $256$-QAM, $\sigma_{\Delta}^{2}=0.5 \times 10^{-3} \text{rad}^{2}$. }\label{fig:ekf_comp_256}
\end{center}
\end{figure}
In the EKF, the soft symbol is treated as the true value of transmitted symbols in order to compute the phase noise estimate $\hat{\phi}_{k}$ and its variance. Iterations between the estimator and detector continue till a maximum number of iterations has been reached (stopping criterion). After the stopping criterion is reached, the detector computes the final hard decisions as
\begin{equation}
 \hat{{s}}_{k} = \underset{{s_{k}} \in{\mathcal{C}}}{\textrm{argmax}}\: P_{\text{app}}(s_{k}).\nonumber
\end{equation}

For simulation results demonstrated in  Fig. \ref{fig:ekf_comp_16}, data is drawn uniformly from a 16-QAM constellation, and transmitted in form of frames that are $10000$ symbols long with 5 pilot symbols being transmitted at the beginning of each frame. Further a pilot symbol is transmitted every 15 data symbols resulting in an overall pilot density of $6.5\%$. The variance of the innovation component is set to $\sigma_{\Delta}^{2}=10^{-2} \text{rad}^{2}$. For the results in Fig. \ref{fig:ekf_comp_256}, data is uniformly drawn from a $256$-QAM constellation, and a pilot density of $6.5\%$ is used. The innovation variance is set to $\sigma_{\Delta}^{2}=0.5 \times 10^{-3} \text{rad}^{2}$. The maximum number of iterations between the estimator and the detector in a time instant $k$ is fixed to $3$ for both constellations.

From Figs. \ref{fig:ekf_comp_16} and \ref{fig:ekf_comp_256}, the following observations can be made about the performance of all detectors that work in conjunction with an EKF: GAP-D and FOS-D perform significantly better than all the other detectors considered as the SNR increases. In fact, for both constellations, GAP-D outperforms FOS-D slightly by about $1$ dB at high SNR. COL-D performs better than all the detectors in the low SNR regime owing to large sensitivity of the EKF to decision feedback errors that are rampant at low SNRs. However, with increase in SNR, the decision feedback errors diminish and the gains from using GAP-D and FOS-D become prominent. SOM-D and VB-D performs better than EUC-D at lower SNRs, but they converge to the performance of EUC-D at high SNRs. This is because the estimation error variance decreases with increase in SNR, as a result of which the term associated with the variance of the phase error in \refeq{eq:sum_moments_2} and \refeq{eq:ml_vb} becomes significantly small.

\begin{figure}[!ht]
\begin{center}
\includegraphics[width = 3.5in, keepaspectratio=true]{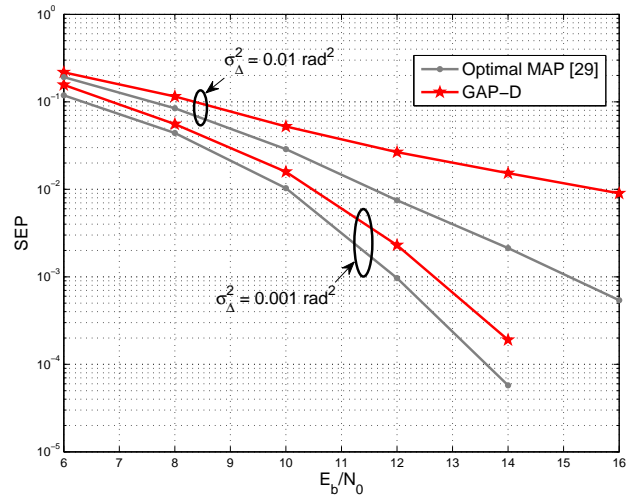}
\caption{Comparison of SEP performance between the GAP-D and optimal MAP \cite{sch81} for $16$-QAM, $\sigma_{\Delta}^{2}= 1  \times 10^{-3} \text{ rad}^{2}, 1 \times 10^{-2} \text{ rad}^{2}$. }\label{fig:sep_comp_opt}
\end{center}
\end{figure}

Now, we compare the performance of GAP-D with an (approximate) optimal joint symbol-phase MAP estimator that have been proposed based on Viterbi algorithm in \cite{sch81} and the BCJR algorithm in \cite{sha98}. We implement this joint estimator using SPA as in \cite{cola05}, and let the phase to assume $D$ discrete values: $\phi_{k} = \{0,2\pi/D,\ldots,2\pi(D-1)/D\}$. Note that the accuracy of the algorithm increases with increase in $D$. We set $D = 8C$ as in \cite{sha98}, where $C$ is the size of the constellation. For complexity reasons, we present the performance of the optimal MAP only for $16-$QAM in Fig. \ref{fig:sep_comp_opt} for different values of $\sigma_{\Delta}^{2}=10^{-2} \text{ rad}^{2}, 10^{-3} \text{ rad}^{2}$. We observe that the performance gap between GAP-D and the optimal MAP is around $1$ dB for $\sigma_{\Delta}^{2}=10^{-3} \text{ rad}^{2}$. This gap significantly increases for $\sigma_{\Delta}^{2}=10^{-2} \text{ rad}^{2}$ that results from the unreliable feedback from the detector to the EKF due to high phase noise. Note that it is possible to extend the premise of the GAP-D (with the estimator) to the case where the phase error PDF is multi-modal.

\section{Conclusions}
\label{sec:conc}
In this work, we derived a joint amplitude-phase detector (GAP-D) for detecting data in the presence of phase error that is Gaussian distributed, which has an intuitive and simple analytical form. We compared the SEP performance of GAP-D with other metrics/detectors available in literature, and observed that GAP-D and FOS-D can achieve significant SEP gains with respect to other existing detectors from literature such as COL-D, EUC-D, VB-D. The GAP-D performs slightly better than FOS-D in terms of SEP for some of the scenarios evaluated. To sum up, significant performance gains can be achieved in terms of SEP when the PDF of phase error is incorporated in the detector for any constellations (especially higher order constellations).

Additionally, we showed that the ML data detector for symbol by symbol detection in the presence of phase noise can be formulated as a weighted sum of central moments of the posteriori PDF of phase noise. We approximated the optimal rule by truncating it to two terms (SOM-D) and observed that this approximation renders SEP gains with respect to EUC-D and VB-D for medium/high phase noise variance and low SNR. Then based on GAP-D, we analytically derived SEP and error floor for arbitrary constellations (which is not possible using FOS-D \cite{kam09}). This was shown to be a tight upper bound on the SEP performance for all values of SNR considered. Also it was inferred that error floor in the presence of phase noise arises due to equal energy points in the signal constellation that are not separated by large angular distances.

The simpler analytical form of GAP-D and its SEP characterization pave way for interesting research directions. Analysis of probability of error derived helps derive constellations that are optimal in the SEP sense, which is an ongoing research endeavor. Using the probability of pairwise error event, we can analytically derive the mutual information of a phase noise channel that uses GAP-D (for e.g., codes with hard-decision decoding). The detectors considered in this paper, which are also referred to as soft metrics, can be used for computing branch metrics for decoding in coded systems, which is also an ongoing effort.

\bibliographystyle{IEEEbib}

\end{document}